\documentclass[a4paper]{jpconf}
\usepackage{graphicx,epsfig}
\begin{document}
\title{Nuclear effects in neutrino-nucleus interactions}

\author{Maria B. Barbaro}

\address{University of Torino and INFN,  Via Giuria 1, I-10125 Torino, Italy}

\ead{Barbaro@to.infn.it}

\begin{abstract}
An accurate description of the nuclear response functions for neutrino 
scattering in the Gev region is essential for the interpretation of present 
and future neutrino oscillation experiments.
Due to the close similarity of electromagnetic and weak scattering processes, 
we will review the status 
of the scaling approach and of relativistic modeling for the
inclusive electron scattering response functions in the quasielastic 
and $\Delta$-resonance regions.
In particular, recent studies have been focused on scaling violations 
and the degree to which these imply modifications of existing 
predictions for neutrino reactions.
We will discuss sources and magnitude of such violations,
emphasizing similarities and differences between electron and  
neutrino reactions.
\end{abstract}

\section{Introduction}

The motivation of the recent interest in neutrino-nucleus interactions
is twofold.
One one hand ongoing and future high precision neutrino oscillation experiments 
increasingly rely on our knowledge of neutrino-nucleus 
interactions in both Charged Current (CC) processes
of the type
$\nu_l+A \to l+N+(A-1)$,
where a virtual $W$ boson is exchanged, and Neutral Current (NC) reactions
$\nu_l+A \to \nu_l^\prime+N+(A-1)$,
where  a $Z$ boson is exchanged.
Typical neutrino energies involved in these experiments are of a few GeV.
In this region a large contribution to the cross section comes from 
quasielastic and resonance processes, where the nuclear dynamics is known
to play an important role.
Since many electron scattering data exist at these kinematics (see 
Ref.~\cite{Benhar:2006er} for a recent review), we should be able to use them
to predict neutrino cross sections. As we will show, the superscaling 
properties of inclusive $(e,e^\prime)$ data represent a bridge between
electron and neutrino scattering reactions off nuclei.

A second motivation consists in the fact that
neutrinos can reveal informations on the nuclear structure 
complementary to electrons
and they can be used to probe the internal
structure of the nucleon (in particular the axial mass and the
strange form factors), providing the nuclear content of the problem is
under control.

In this contribution we will first review the phenomenon
of superscaling in inclusive electron-nucleus 
scattering~\cite{West:1974ua,Alberico:1988bv,Day:1990mf,Donnelly:1998xg} and then show
some predictions for $\nu$-$A$ cross sections based on the
SuperScaling Approximation (``SuSA'')~\cite{Amaro:2004bs,Amaro:2006pr}.
We will then discuss the microscopic origin of the superscaling function,
with particular focus on two models, the
Relativistic Mean Field (RMF)~\cite{Caballero:2005sj}
and a BCS-like correlated Fermi gas~\cite{Barbaro:2008zv}, which seem to 
correctly reproduce the shape of the $(e,e^\prime)$ experimental data,
at variance with other relativistic models.
Finally we will address the issue of scaling violations, in particular
the ones connected to meson-exchange currents.

\section{Superscaling}

The experimental quasielastic $(e,e^\prime)$ cross sections display the so-called
``first kind scaling''~\cite{Donnelly:1998xg} behaviour: at high momentum transfer 
($q$ larger than about 0.5 GeV/c) the reduced quasielastic cross section
\begin{equation}
F = \frac{d\sigma}{d\Omega d\omega}/\left(Z\sigma_{ep}+
N\sigma_{en}\right)
\label{F}
\end{equation}
does not depend on two variables ($q$, $\omega$) but only on one combination
of them, called the scaling variable and indicated by 
$y(q, \omega)$ in a non-relativistic scheme and $\psi(q, \omega)$  in a
relativistic context. These two scaling variables represent respectively the
minimum momentum $(-y)$ or kinetic energy $(m_{_N}\psi)$ that a nucleon
inside the target nucleus must have in order to participate to the reaction.
The conditions $y=0$ and $\psi=0$ correspond, in the non-relativistic and 
relativistic case respectively, the quasielastic peak (QEP).
In Ref.~\cite{Day:1990mf} it was shown that if the reduced function (\ref{F}) 
corresponding to different kinematical conditions is plotted as a function of
$y$ it fulfills with good accuracy first kind scaling in the so-called scaling region
$y<0$, whereas at $y>0$ scaling is broken.

Second kind scaling consists instead in the observation that
for different target nuclei
the reduced cross section scales as the inverse of the Fermi momentum $k_F$,
so that the so-called {\em superscaling function}
$f =k_F\times F$ is independent of the specific nucleus.
This has been illustrated in Ref.~\cite{Donnelly:1998xg}, where it has been shown that 
scaling of second kind is excellent in the scaling region and slighthly
violated at the right of the QEP.

If first and second kind scaling are simultaneously respected the cross sections are
said to ``superscale''. 
A closer inspection of the Rosenbluth-separated longitudinal and transverse 
response functions shows that the 
scaling violations observed at $\psi>0$ mainly reside in the transverse 
channel, essentially due to the $\Delta$ resonance contribution.

Before proceeding, let us briefly remind the basic formalism for inclusive electron scattering, 
focussing on the longitudinal 
channel where superscaling is seen to work better.
In plane wave impulse approximation the longitudinal 
quasielastic response function can be expressed as
\begin{equation}
  R_{L}(q,\omega)=\frac{2\pi m_{_N}^2}{q} \int\int_\Sigma dp\,d{\cal E}\,
S(p,{\cal E})
{\cal R}_{L}(q,\omega;p,{\cal E}) ,
\end{equation}
where $\Sigma(q,\omega)$ 
is the kinematically allowed region in the $({\cal E},p)$ plane,
$\cal E$ being the excitation energy of the residual nucleus and $p$ its 
momentum, $S(p,{\cal E})$ is the
nuclear spectral function and ${\cal R}_{L}$ the single nucleon response,
which depends on the electric and magnetic nucleon's form factors.

Assuming that the single nucleon response ${\cal R}_{L}$ is smoothly varying in
the $({\cal E},p)$-plane and can be factored out of the integral one gets
\begin{equation}
R_L(q,\omega)=
{\cal R}_L(q,\omega;p=p_{\rm min},{\cal E}=0) \times 
F_L(q,\omega)
\end{equation}
where
\begin{equation}
  F_L(q,\omega) = \frac{2\pi m_{_N}^2}{q} \int\int_{\Sigma(q,\omega)} dp\, d{\cal E} 
S(p,{\cal E})
\end{equation}
is the longitudinal scaling function.
Note that the above factorization can be only approximately true in a 
relativistic context and it is violated to some degree due to off-shell effects 
and breaking of independent particle models.

\begin{figure}
\begin{center}
\label{fig:scal}
\includegraphics[height=5.cm]{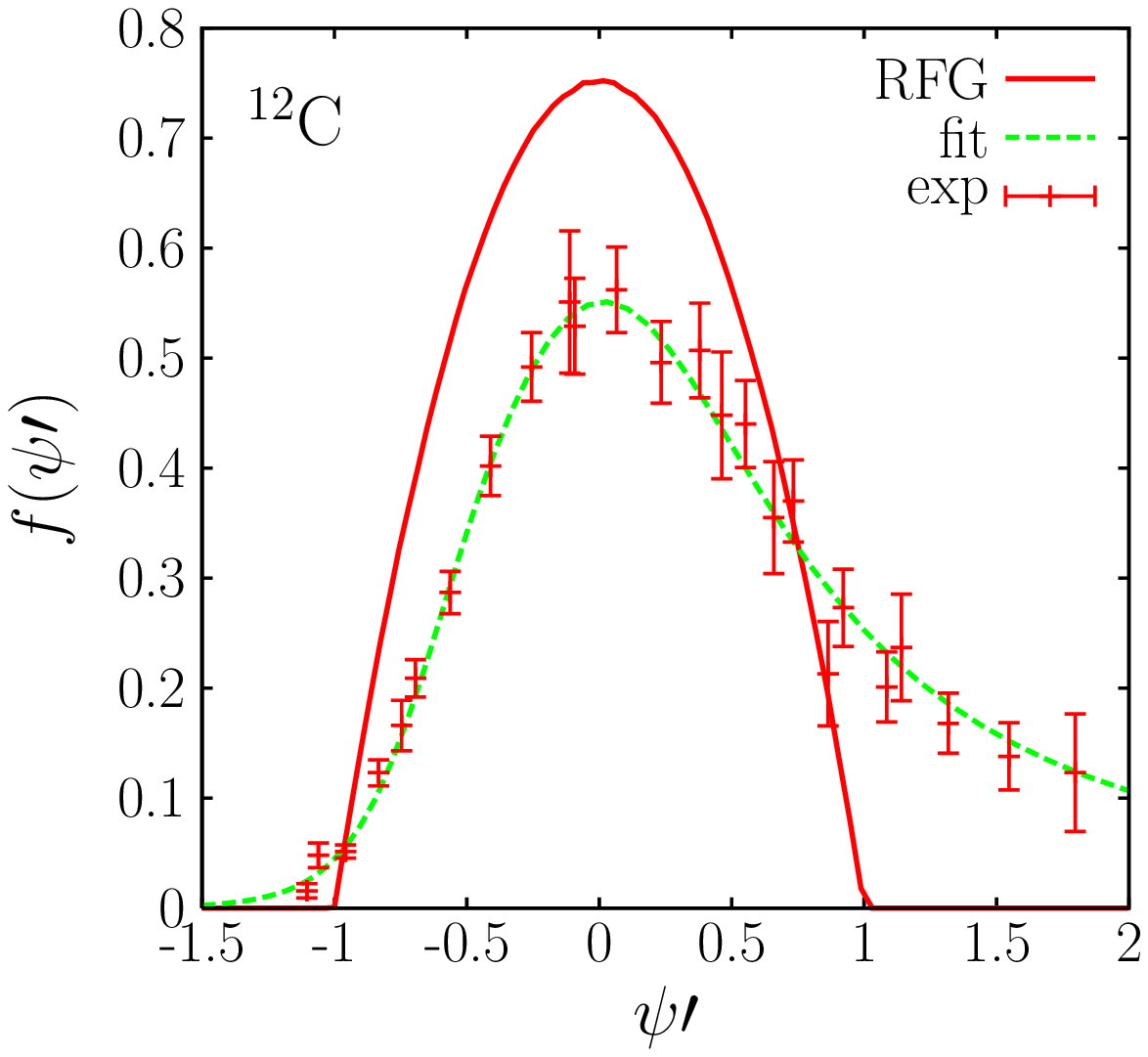}\ \ \  %
\includegraphics[scale=0.38,clip,angle=0]{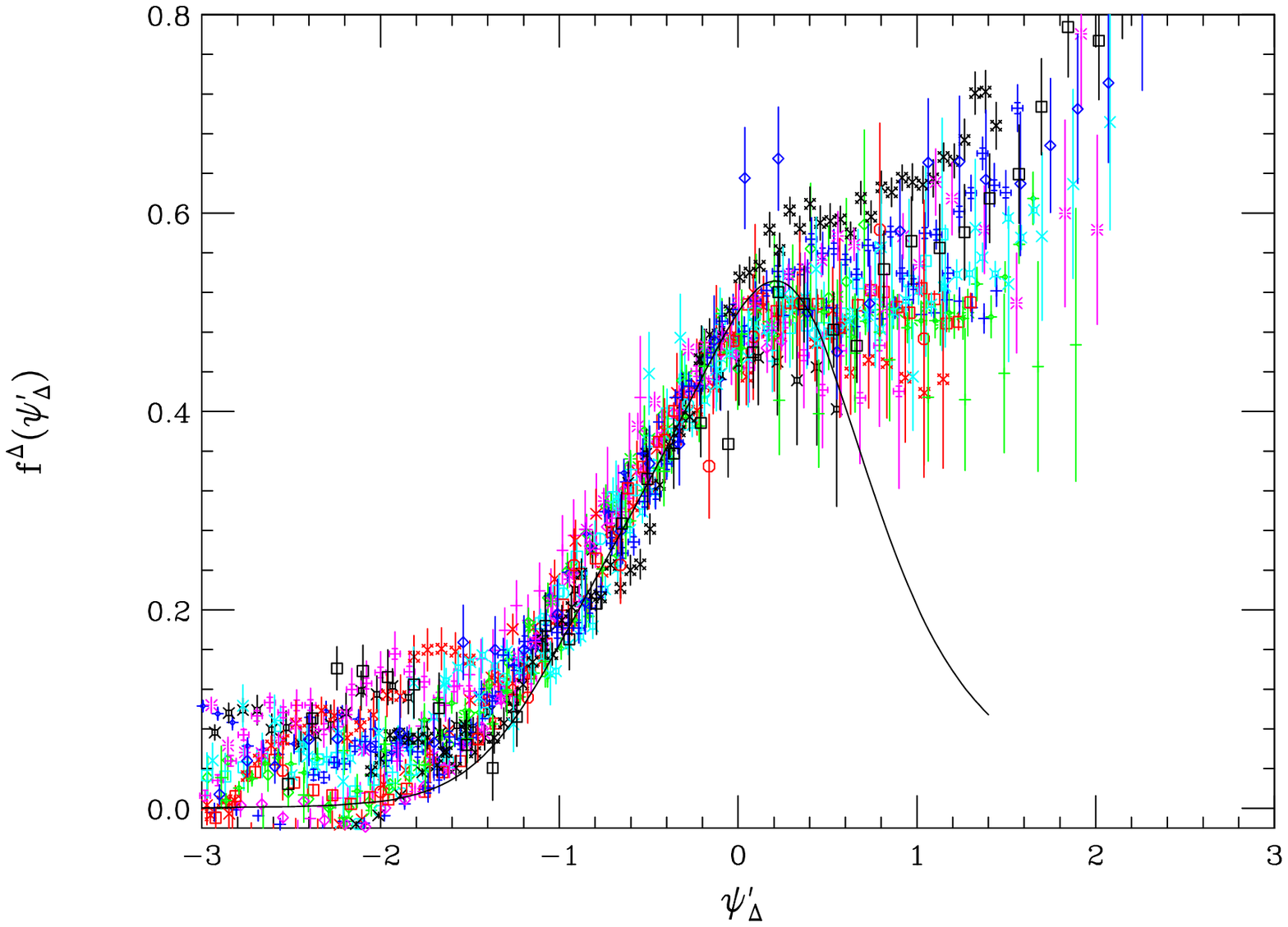}
\caption{The experimental superscaling function in the quasielastic 
(left panel)~\cite{Jourdan:1996ut}
and $\Delta$ resonace (right panel) regions.}
\end{center}
\end{figure}

The simplest model where one can perform exact relativistic calculations is the 
Relativistic Fermi Gas (RFG), namely an independent-particle model in which on-shell 
nucleons are described by Dirac spinors. In spite of its simplicity the RFG model retains some 
fundamental properties: it is Lorentz covariant and gauge invariant and is therefore a 
good starting point for more sophisticated calculations.
The RFG spectral function is
$
S^{RFG}(p,{\cal E}) =  
\theta(k_F-p) \delta({\cal  E}-T_F+T_{\bf p})
$
and the corresponding scaling function is
\begin{equation}
F_L(q,\omega) = 
\left(\frac{m_{_N}}{q}\right) 
\left(\frac{1}{k_F}\right) 
\frac{3}{4}
\left(1-\psi^2\right)
\theta\left(1-\psi^2\right)
\label{eq:FRFG}
\end{equation}
where the parabola 
$\frac{3}{4}\left(1-\psi^2\right)\theta\left(1-\psi^2\right)\equiv f(\psi)$ 
is the superscaling function and $\psi\equiv\psi(q,\omega)$ is the RFG scaling variable.
The RFG model exactly superscales, as the data do in the so-called scaling region.
However the comparison with the averaged data, illustrated in the left panel of Fig.~\ref{fig:scal},
shows that the corresponding superscaling function is very different from the experimental one
and in particular it fails the reproduce the long tail displayed by the data at large positive $\psi$.
However a simple phenomenological approach can be taken for describing the QE region: a fit 
of the experimental superscaling function can be extracted from the data and plugged into
the RFG formulas in place of the parabola (\ref{eq:FRFG}). Note that the fit involves
only four parameters for describing {\em all} kinematics and {\em all} nuclei.
Moreover, it represents a strong constraint on  nuclear models, since any reasonable model
used to describe the quasielastic region must reproduce this shape.

The scaling analysis has been recently extended to the $\Delta$ resonance 
region~\cite{Amaro:2004bs}.
In order to isolate the contributions in this region,
the impulsive contributions arising from quasielastic $eN$ scattering, calculated
assuming superscaling is valid, has been 
removed from the experimental total cross section and the result has been
divided by the elementary $N\to\Delta$ cross section, thus leading to
a new superscaling function
\begin{equation}
f_\Delta (q,\psi_\Delta)
\equiv k_F \frac{ \left[d^2\sigma / d\Omega_e d\omega\right]_{\Delta} }
    { \sigma _{M}\left[ v_{L}G_{L}^{\Delta }+v_{T}G_{T}^{\Delta
}\right] } ~.
\end{equation}
When the latter is plotted against the appropriate scaling variable
$\psi_\Delta\equiv\psi(q\rho,\omega\rho)$,
which accounts for the different kinematics through the inelasticty paramater
$\rho=1+(m_{_\Delta}^2-m_{_N}^2)/(4|Q^2|)$, the result
shown in the right panel of Fig.~\ref{fig:scal} is obtained.
Obviously this approach can work only at $\psi'_\Delta < 0$, since at $\psi'_\Delta > 0$
other resonances and the tail of DIS contribute.

The two above superscaling functions can then be used to reconstruct the 
$(e,e^\prime)$ cross sections in the QE and $\Delta$ regions, giving a satisfactory description
of the data, as shown in Ref.~\cite{Amaro:2004bs}.

\section{Predictions for neutrino scattering: the SuSA approach}

We shall now illustrate how the superscaling properties of inclusive electron scattering
can be exploited to predict neutrino cross sections off nuclei.

Charged current neutrino-nucleus scattering can be easily linked to inclusive $(e,e^\prime)$
reactions by replacing the virtual photon by a wirtual $W$-boson. The main
difference is that in the neutrino case, beyond the longitudinal and transverse response functions, a third
response is involved, arising from the axial component of the weak current.
In the RFG model the three response functions turn out to be 
\begin{equation}
{\widetilde R_i}= 
 {\widetilde G_i}\times f(\psi)~,
\label{eq:wresp}
\end{equation}
where the functions ${\widetilde G_i}$ depend on the weak nucleon form factors and the
information about the nuclear dynamics is carried by the superscaling function $f(\psi)$ .
The basic assumption of the SuSA approximation is that the nuclear dynamics governing 
the neutrino-nucleus process is the same entering the electron scattering reactions and
therefore the corresponding cross sections in the QE and $\Delta$ regions 
can be obtained by simply replacing
in (\ref{eq:wresp}) the RFG superscaling function with the corresponding
phenomenological fits:
$f_{RFG} \Rightarrow f_{QE,\Delta}$.

\begin{figure}[hbt]
\begin{center}
\label{fig:xx}
\includegraphics[scale=0.4,clip]{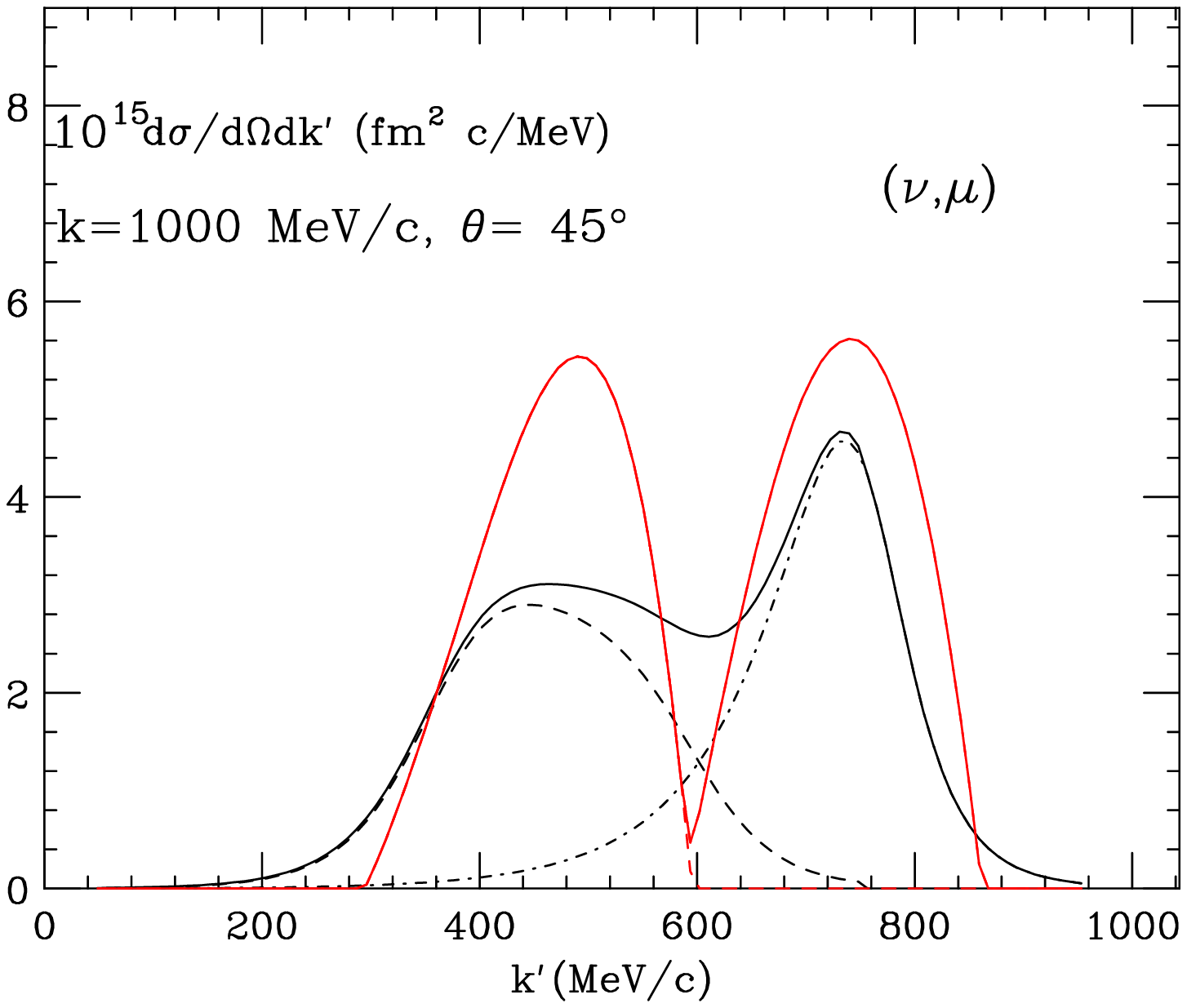}%
\ \ \ \ \includegraphics[scale=0.55,clip]{integratedcs.epsi}
\caption{Left: CC neutrino reaction cross sections on $^{12}C$
versus the outgoing muon momentum $k^\prime$, showing a comparison of  
the SuSA  results (black online) with the RFG (red onleine); the separate
QE and $\Delta$ contributions are shown with dashed lines.
Right: integrated QE cross section
versus the neutrino energy evaluated in different models.}
\end{center} 
\end{figure}

In the left panel of Fig.~\ref{fig:xx} the double differential cross section in the QE 
and $\Delta$ regions is shown for 
(see Refs.~\cite{Amaro:2004bs} for more results):
it clearly appears that in both regions the SuSA result lies significantly
lower and extends over a wider range in $k'$ than the RFG.
In the right panel the SuSA fully integrated quasielastic cross section~\cite{Amaro:2006tf} on
$^{12}C$ is shown as a function of the neutrino energy and compared not only with the RFG but also
with other relativistic models: the relativistic mean field (RMF) model, the
relativistic plane wave impulse approximation (RPWIA) and a semi-relativistic shell model 
without (SRWS) or with (SRWS-tot) inclusion of discrete states of $^{12}N$. Note that the RMF
yields results which are very similar to the SuSA prediction, whereas the other models
are closer to the RFG.
The comparison of these results with the recent MiniBooNE data~\cite{Katori:2009du}, which lie
above the RFG and seem to point to a very high value of the axial mass ($M_A c^2=1.35$ GeV
in place of the usual value 1.03, which we are using in our calculations), is somehow puzzling and
still needs to be carefully explored.

\begin{center}
\begin{figure}
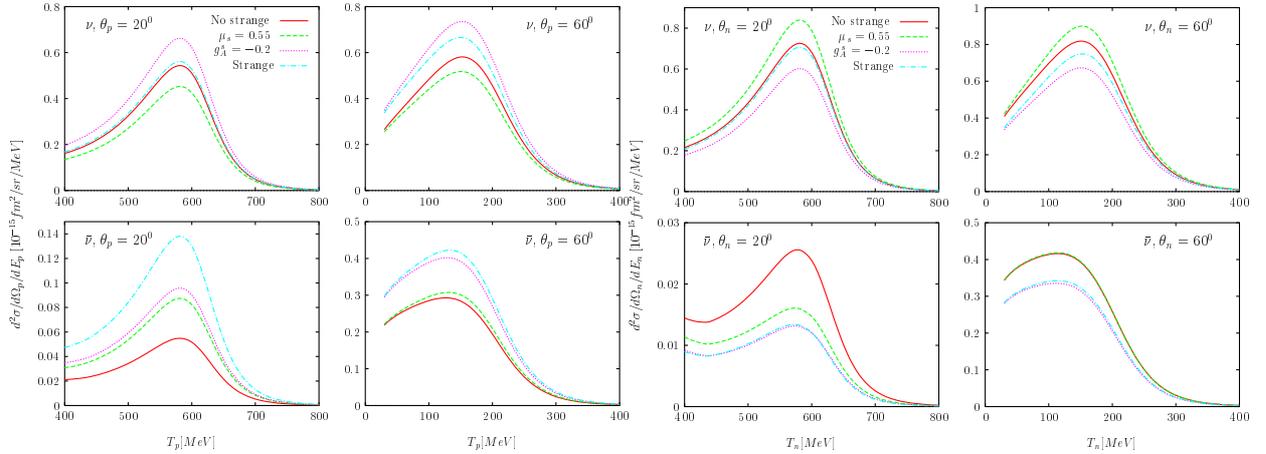

\label{fig:strpn}
\includegraphics[scale=0.4]{fig4_p.epsi}%
\includegraphics[scale=0.4]{fig4_n.epsi}
\caption{NC neutrino (upper panels) and antineutrino (lower panels) 
cross sections for proton (left) and neutron (right) knockout. The effect of strangeness is
shown in the various curves.}
\end{figure}
\end{center}

Concerning neutral current neutrino reactions,
it has been proved in Ref.~\cite{Amaro:2006pr} that, although the kinematics
is not the same of $(e,e^\prime)$ reactions because the outgoing detected particle is
the knocked nucleon instead of the lepton,
the SuSA approach can still be used in the QE channel to predict cross sections.
The result is qualitatively similar to the CC case, namely the SuSA predictions
are lower the RFG and the response estends to a wider region. 
As anticipated an interesting feature of NC reactions is their sensitivity to the
strangeness content of the nucleon, which allows in principle to extract informations
on the strange hadronic form factors by different combinations of neutrino
and antineutrino scattering off protons and 
neutrons~\cite{Barbaro:1996vd,Alberico:1997rm,Meucci:2006ir}.
This is illustrated in the examples of Figs.~\ref{fig:strpn}, where it is seen that
for proton knockout (left panels) the magnetic strangeness decreases (increases) 
the $\nu$ ($\overline\nu$) 
cross section, while the 
axial strangeness increases both $\nu$ and $\overline\nu$ cross sections.
For neutron knockout (right panels) and $\nu$ scattering 
the situation is reversed with respect to p-knockout, while
for $\overline\nu$ scattering the effect of magnetic strangeness is very
small due to cancellation between the vector and axial-vector responses.

\section{Microscopic origin of the superscaling function}

In this section we will briefly illustrate the results of two relativistic models
which, although sensibly different, seem to reproduce the basic properties of
the experimental quasielastic superscaling function, namely a scaling behaviour of both first and 
second kind and a long high energy tail.

\begin{figure}
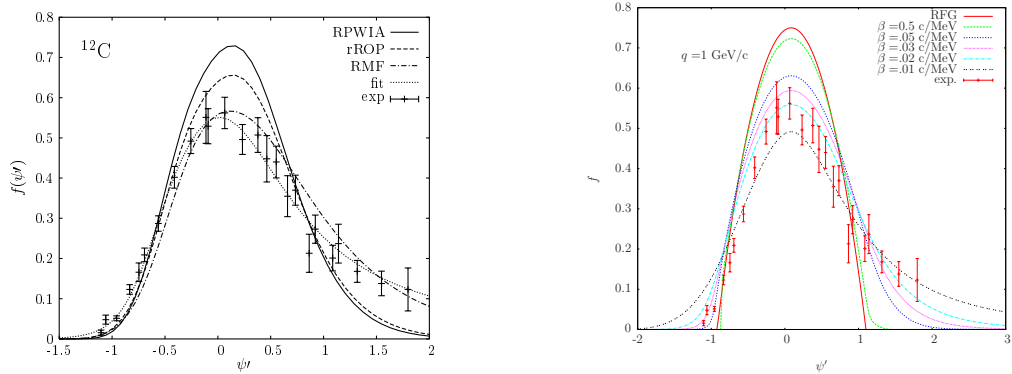

\centering
\includegraphics[scale=0.4,clip]{RMF.epsi}\hspace{2.cm}%
\includegraphics[scale=0.4,clip]{Flbcs.epsi} 
\caption{Left: QE superscaling function $f$ evaluated in the RMF model compared with other 
models (see text); right: $f$ in the BCS-like model for 
different values of the parameter $\beta$. }
\label{fig:models}
\end{figure}

The first model is the Relativistic Mean Field model,
based on a Lagrangian containing strong scalar and vector potentials 
(corresponding to $\sigma$, $\omega$ and $\rho$ mesons) and describing the bound nucleon states 
as self-consistent Dirac-Hartree solutions. The final state interactions (FSI) 
are described in distorted wave impulse approximation using the same
relativistic potentials~\cite{Caballero:2005sj}. 
As can be seen in Fig.~\ref{fig:models} (left panel) 
the RMF model provides the correct amount of asymmetry, unlike other relativistic models, also
shown in the same plot. These correspond to the absence of FSI (RPWIA) and to a treatment of
FSI through a real optical potential (rROP). It clearly emerges from the results that 
not only relativity, but also a consistent treatment of the initial and final states are required
in order to get the observed asymmetry of the scaling function.
Morevoer, as shown in Ref.~\cite{Caballero:2005sj}, 
the model fulfills superscaling with good accuracy.

A completely different model, which we will refer to as ``BCS-like'', has been shown
to give very similar results~\cite{Barbaro:2008zv}, as illustrated in Fig.~\ref{fig:models} (right
panel). It is a simple covariant extension of RFG which accounts for NN correlations
through the following wave functions
\begin{equation}
  |\Phi>=\prod_k(u_k+v_k a^\dagger_{k\uparrow}a^\dagger_{-k\downarrow})|0>
\ \ \ \ \mbox{and}\ \ \ \ 
  |D(p)>=\frac{1}{|v_p|}\,a_{p\uparrow}
\prod_k[u_k+v_k
   a^\dagger_{k\uparrow}a^\dagger_{ -k\downarrow}]|0>
\end{equation}
for the initial and final states, respectively.
The $v_k$-coefficients are taken phenomenologically as Fermi-Dirac distributions
with one free parameter $\beta$, controlling the tail of the momentum distribution $v_k^2$.
The associated superscaling function
$f$ turns out to be asymmetric in $\psi$, the asymmetry increasing with the value
of $\beta$, and to reproduce the data for values of $\beta\simeq 0.02$ c/MeV.
The superscaling properties of the BCS-like 
model have been explored in Ref.~\cite{Barbaro:2008zv}, where
it was shown that both kinds of scaling are good at the QEP and broken to some degree at the right 
and left of it.

\section{Scaling Violations}

At low momentum and energy transfer (which can be defined approximately by  
$q < 400$ MeV/c and $\omega<50$ MeV) the scaling approach is bound to fail because 
collective effects, like giant resonances, become important.
However even at higher energies scaling violations are observed and it is very interesting
to understand their origin.

\begin{center}
\begin{figure}
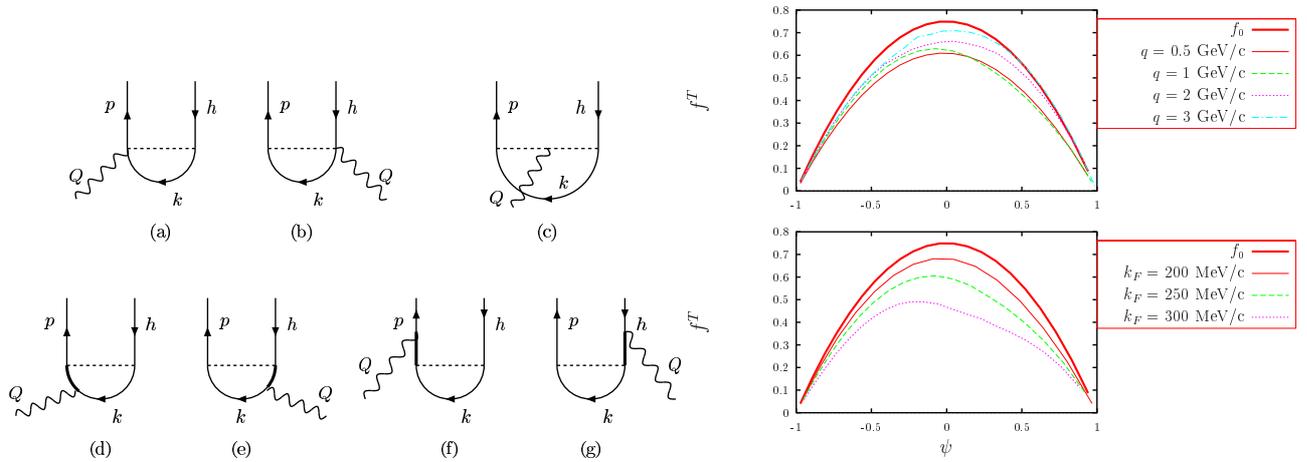

\includegraphics[height=5.cm]{phdiag.epsi}
\includegraphics[height=6.cm]{figdelta_scaling.epsi}
\caption{1p-1h MEC diagrams and results for $f$ in the RFG ($f_0$) and including the 
$\Delta$-MEC diagram for different values of $q$ (upper panel) and $k_F$ (lower panel).}
\label{fig:MEC}
\end{figure}
\end{center}

Among other possible contributions, like {\em e.g.} short range correlations, meson-exchange 
currents (MEC) can certainly play a significant role in breaking
superscaling in the transverse channel.
The MEC, being two-body currents, contribute to both the 1p-1h and the 2p-2h sectors of the 
response and their calculation is far from trivial in a relativistic context.
Moreover in order to preserve gauge invariance, in parallel with the pure MEC diagrams, where
the virtual photon attaches to the exchanged meson, it is
necessary to take into account the corresponding correlation diagrams, where the photon attaches
to the nucleon. Up to now an exact gauge invariant relativistic calculation has been performed
only in the RFG system and in the 1p-1h channel~\cite{Amaro:1998ta,Amaro:2002mj,Amaro:2003yd}.
Here the diagrams are the ones shown in Fig.~\ref{fig:MEC}, where thick lines in 
(d)-(g) represent either a nucleon or a $\Delta$.
It turns out that the diagrams involving the $\Delta$ give the dominat contribution and they
break both kinds of scaling, as illustrated in the right panel of Fig.~\ref{fig:MEC}.
Moreover a recent calculation~\cite{Amaro:2009dd} performed in a semi-relativistic shell model 
essentially equivalent to the RMF with FSI has shown that the MEC contribution to the high 
energy tail of the response can become very large due to the dynamical nature of the pion 
propagator, an effect which is emphasized by the presence of FSI.

In the 2p-2h sector the pure MEC have been calculated on the RFG basis~\cite{De Pace:2003xu} and
shown to give a substantial contribution at high negative and positive values of $\psi$, 
but a full gauge invariant calculation is still missing.

\section{Conclusions} 

We have shown how the superscaling properties of $(e,e')$ data 
can be exploited to predict charged and neutral
current neutrino scattering cross sections
and we have proposed two microscopic relativistic models, the 
RMF model and the BCS-like correlated Fermi gas, which are capable of reproducing the 
experimental superscaling function, in particular its long high energy tail.

However, in order to assess the applicability of the SuSA approach and eventually improve
the method we need not only to justify
the microscopic origin of superscaling but also to understand its violations, in particular the
ones associated with meson exchange currents.
This has been partially accomplished in recent work~\cite{Maieron:2009an}
 and more effort will be done in this
direction in order to provide a consistent description of the nuclear dynamics in the
quasielastic, resonance and deep inelastic regions. 

\section*{Acknowledgments}
The work presented here is the result of collaborations with J.E. Amaro, J.A. Caballero, R. Cenni,
T.W. Donnelly, C. Maieron, A. Molinari, I. Sick and J.M. Udias.

\end{document}